____________________________________________________________________
\documentstyle[12pt]{article}
\input psfig.sty
\begin{document}

\title{Schr\"{o}dinger Equation \\
with the Potential $V(r)=a r^2+b r^{-4}+c r^{-6}$}

\author{Shi-Hai Dong\thanks{Electronic address: DONGSH@BEPC4.IHEP.AC.CN}\\
{\scriptsize Institute of High Energy Physics, P. O. Box 918(4), 
Beijing 100039, People's Republic of China}\\
\\
Xi-Wen Hou \\
{\scriptsize Institute of High Energy Physics, 
P. O. Box 918(4), Beijing 100039, and}\\
{\scriptsize Department of Physics, Hubei University, 
Wuhan 430062, People's Republic of China} \\
\\
Zhong-Qi Ma\\
{\scriptsize China Center for Advanced Science and Technology
(World Laboratory), P. O. Box 8730, Beijing 100080}\\
{\scriptsize  and Institute of High Energy Physics, P. O. Box 918(4), 
Beijing 100039, People's Republic of China}}

\date{}

\maketitle

\begin{abstract}
By making use of an ${\it ansatz}$ for the eigenfunction, 
we obtain the exact solutions to the Schr\"{o}dinger equation 
with the anharmonic potential, $V(r)=a r^2+b r^{-4}+c r^{-6}$,
both in three dimensions and in two dimensions, where 
the parameters $a$, $b$, and $c$ in the potential satisfy 
some constraints.



\end{abstract}

\newpage

\begin{center}
{\large 1. Introduction}\\
\end {center}

The exact solutions to the fundamental dynamical equations
play crucial roles in physics. It is well-known that the 
exact solutions to the Schr\"{o}dinger equation have been
obtained only for a few potentials, and some approximate methods 
are frequently applied to arrive at the approximate solutions.  
In recent years, the higher order anharmonic potentials have 
drawn more attentions of physicists and mathematicians 
in order to partly understand a few newly discovered
phenomena, such as the structural phase transitions [1], 
the polaron formation in solids [2], and the concept
of false vacuo in field theory [3]. Interest in these 
anharmonic oscillator-like interactions comes from the fact 
that the study of the relevant Schr\"{o}dinger equation, for 
example, in the atomic and molecular physics, provides us with 
insight into the physical problem in question.

For the Schr\"{o}dinger equation ($\hbar=2m=1$ for convenience)
$$-\nabla^{2} \psi +V(r) \psi =E \psi, \eqno (1) $$ 

\noindent
with the potential
$$V(r)=a r^2+b r^{-4}+c r^{-6},~~~~~ a>0, ~~c>0, \eqno (2) $$

\noindent
let
$$\psi(r,\theta, \varphi)=r^{-1} R_{\ell}(r) Y_{\ell m}(\theta, \varphi), 
\eqno (3) $$  

\noindent
where $\ell$ and $E$ denote the angular momentum and the 
energy, respectively, and the radial wave function 
$R_{\ell}(r)$ satisfies
$$\displaystyle {d^{2} R_{\ell}(r) \over dr^{2} }
+\left[E-V(r)-\displaystyle {\ell(\ell+1) \over r^{2}} \right] 
R_{\ell}(r)=0. \eqno (4) $$

Znojil [4,5] converted Eq. (4) into a difference
equation in terms of a Laurent-series ansatz for the radial
function
$$R_{\ell}(r)=N_{0}r^{\kappa}\exp [-(\sqrt{a}r^{2}+\sqrt{c}
r^{-2})/2] \displaystyle \sum_{m=-M}^{N}~h_{m}r^{2m}. \eqno (5) $$

\noindent
He defined the continued fraction solutions to accelerate the
convergence of the series, and obtained the solutions
for the ground state and the first excited state.

Kaushal and Parashar highly simplified the ansatz for
calculating those solutions 
$$R_{0}(r)=N_{0}r^{\kappa_{0}}\exp [-(\sqrt{a}r^{2}+\sqrt{c} r^{-2})/2]
,~~~~~\kappa_{0}=(b+3\sqrt{c})/(2\sqrt{c}),  \eqno (6) $$

\noindent
for the ground state [6], and
$$R_{0}(r)=N_{1}r^{\kappa_{1}}\left(1+\beta r^{2}+\gamma r^{-2}
\right)\exp [-(\sqrt{a}r^{2}+\sqrt{c} r^{-2})/2] , \eqno (7) $$

\noindent
for the first excited state [7]. By this ansatz, the
parameters in the potential (2) have to satisfy two constraints:
$$\left(2\sqrt{c}+b\right)^{2}=c\left[(2\ell+1)^{2}+8\sqrt{ac}\right], 
\eqno (8) $$

\noindent
and
$$\begin{array}{l}
\eta_{\ell}\left[(\eta_{\ell}-4)^{2}-4(2\kappa_{1}-1)^{2}\right]
=64\sqrt{ac}(\eta_{\ell}-4), \\
\kappa_{1}=(b+7\sqrt{c})/(2\sqrt{c}),~~~~~
\eta_{\ell}=\ell(\ell+1)+2\sqrt{ac}-\kappa_{1}^{2}+\kappa_{1}.
\end{array} \eqno (9) $$

\noindent
where there was a sign misprint in [7] (see Eq. (13) in [7]). 
They set the values of the parameters by
$$\ell=0,~~~~~a=1.0,~~~~~c=0.18,~~~~~b=0.04082,
 \eqno (10) $$

\noindent
and found that $\beta=-0.1787$ and $\gamma=0.8485$,
and the energies for the ground state and the first 
excited state were $E_{0}=4.096214$ and $E_{1}=12.09621$, 
respectively. Unfortunately, their parameters given in Eq. (10) 
do not satisfy the second constraint (9), such that the so-called
solution of the first excited state in [7] does not satisfy
Eq. (4). As a matter of fact, they assumed
that the angular momentum $\ell$ is same for both the ground 
state and the first excited state, and that the normalized 
factor $N_{1} \neq 0$, so that they must obtain, as shown 
in Sec. 2 of the present letter, infinite solutions for 
$\beta$ and $\gamma$ if the parameters in the 
potential satisfy the constraints (8) and (9). 

In our viewpoint, Kaushal and Parashar presented a good
idea for studying the Schr\"{o}dinger equation (1) with 
the higher order anharmonic potential (2), but their
calculation was wrong. In the present letter, we 
recalculate the solutions following their idea, and then, 
generalize this method to the two-dimensional Schr\"{o}dinger 
equation because of the wide interest in lower-dimensional 
field theories recently. Besides, with the advent of growth technique 
for the realization of the semiconductor quantum wells, the 
quantum mechanics of low-dimensional systems has become a 
major research field. Almost all of the computational
techniques developed for the three-dimensional 
problems have already been extended to two dimensions.  

This letter is organized as follows. In Sec. 2, we 
recalculate the ground state and the first excited state
of the Schr\"{o}dinger equation with this potential using 
an ${\it ansatz}$ for the eigenfunctions. This method is
applied to two dimensions in Sec.3. The figures for the 
unnormalized radial functions of the solutions are plotted 
in the due sections.

\vspace{4mm}

\begin{center}
{\large 2. Ansatz }
\end{center}

Assume that the radial function in Eq. (3) is
$$R_{\ell}(r)=r^{\kappa}\left(\alpha+\beta r^{2}+\gamma r^{-2}
\right)\exp [-(\sqrt{a}r^{2}+\sqrt{c} r^{-2})/2] , \eqno (11) $$

\noindent
where $\beta=\gamma=0$ and $\kappa=\kappa_{0}$ for the ground 
state, and $\beta \neq 0$, $\gamma \neq 0$ and $\kappa=\kappa_{1}$ 
for the first excited state. Substituting Eq. (11) into Eq. (4), 
we have
$$\begin{array}{rl}
\displaystyle {d^{2} R_{\ell}(r) \over dr^{2} }
&=~\left\{r^{4} a \beta 
+r^{2}[\alpha a-\beta E]
+[-\alpha E+ \beta \ell (\ell+1)+ \gamma a] \right. \\
&~~~+r^{-2}[\alpha \ell (\ell +1)+\beta b-\gamma E] 
+r^{-4}[\alpha b+\beta c+\gamma \ell (\ell +1) ] \\
&~~~\left.+r^{-6}[\alpha c+ \gamma b]
+r^{-8}  \gamma c \right\} 
r^{\kappa} \exp [-(\sqrt{a}r^{2}+\sqrt{c} r^{-2})/2] .
\end{array}  \eqno (12a) $$

\noindent
On the other hand, the derivative of the radial function 
can be calculated directly from Eq. (11),
$$\begin{array}{rl}
\displaystyle {d^{2} R_{\ell}(r) \over dr^{2} }
&=~\left\{r^{4} a \beta 
+r^{2}[a \alpha-\beta \sqrt{a}(2\kappa+5)]\right. \\
&~~~+[- \alpha \sqrt{a} (2\kappa+1)+\beta (2+3\kappa+\kappa^{2}-2\sqrt{ac})
+\gamma a ]  \\
&~~~+r^{-2}[\alpha (\kappa^{2}-\kappa-2\sqrt{ac})
+\beta \sqrt{c} (2\kappa+1)-\gamma \sqrt{a} (2\kappa-3)]\\
&~~~+r^{-4}[\alpha \sqrt{c}(2\kappa -3)+\beta c 
+\gamma(6-5\kappa-2\sqrt{ac}+\kappa^{2})]\\
&\left.~~~+r^{-6}[ \alpha c+\gamma \sqrt{c}(2\kappa-7)
+r^{-8} \gamma c\right\} 
r^{\kappa} \exp [-(\sqrt{a}r^{2}+\sqrt{c} r^{-2})/2] .
\end{array} \eqno (12b) $$

\noindent
Comparing the coefficients in the same power of $r$, we obtain
$$\beta [E-\sqrt{a}(2\kappa+5)]=0, \eqno (13a) $$
$$\alpha [E-\sqrt{a}(2\kappa+1)]
=\beta[\ell(\ell+1)+2\sqrt{ac}-\kappa^{2}-3\kappa-2], \eqno (13b) $$
$$\alpha [\ell(\ell+1)+2\sqrt{ac}-\kappa^{2}+\kappa]=
\beta [-b+\sqrt{c}(2\kappa+1)]+\gamma[E-\sqrt{a}(2\kappa-3)], \eqno (13c) $$
$$\alpha [b-\sqrt{c}(2\kappa-3)]=
-\gamma [\ell(\ell+1)+2\sqrt{ac}-\kappa^{2}+5\kappa-6], \eqno (13d) $$
$$\gamma [b-\sqrt{c} (2\kappa-7)]=0. \eqno (13e) $$

For the ground state, $\beta=\gamma=0$ and $\alpha\neq 0$, 
we obtain a constraint (8) and 
$$\kappa_{0}=(3\sqrt{c}+b)/(2\sqrt{c}),~~~~~
E_{0}=\sqrt{a/c}~(b+4\sqrt{c}). \eqno (14) $$

\noindent
For the first excited state, $\beta \neq 0$ and 
$\gamma \neq 0$. From Eqs. (13a) and (13e) we have
$$\kappa_{1}=(7\sqrt{c}+b)/(2\sqrt{c}),~~~~~
E_{1}=\sqrt{a/c}~(b+12\sqrt{c}). \eqno (15) $$

\noindent
It is easy to see from Eqs. (8) and (15) that the right hand
side of Eq. (13d) becomes zero, namely, $\alpha=0$. Since 
Kaushal and Parashar [7] assumed $\alpha \neq 0$, they
must obtain the infinite $\gamma$ if the parameters in 
the potential satisfy two constraints (8) and (9). Now,
we obtain from Eq. (13)
$$\alpha=0,~~~~~~  \gamma=-\sqrt{c/a}~\beta, 
\eqno (16) $$

\noindent
and another constraint 
$$b=-6\sqrt{c}. \eqno (17) $$

\noindent
It is easy to check that the constraints (8) and (17) coincide
with the constraints (8) and (9). 

Setting $\ell=0$ and $a=1.0$ for comparison with Znojil [4] 
and Kaushal-Parashar [7], we obtain
$$\begin{array}{llll}
b=-11.25,~~~~~&\sqrt{c}=1.875,~~~~~&\gamma=-1.875 \beta,~~~~~& \\
\kappa_{0}=-1.5, &\kappa_{1}=0.5,~~~~~ 
&E_{0}=-2,~~~~~ &E_{1}=6. \end{array} \eqno (18) $$

\noindent
Thus, the radial functions $R^{(0)}_{0}(r)$ for the ground state and 
$R^{(1)}_{0}(r)$ for the first excited state are 
$$\begin{array}{l}
R^{(0)}_{0}(r)=N_{0}r^{-1.5}\exp \{-(r^{2}+1.875 r^{-2})/2\},\\
R^{(1)}_{0}(r)=N_{1}r^{-0.5}(r^{2}-1.875 r^{-2})
\exp \{-(r^{2}+1.875 r^{-2})/2\},
\end{array} \eqno (19) $$

\noindent
where the normalized factors are calculated by the normalized
condition:
$$\displaystyle \int_{0}^{\infty}|R_{0}^{(i)}(r)|^{2} dr=1,~~~~~ 
i=0 {\rm ~~and}~~ 1. \eqno (20) $$

\noindent
Without loss of any main property, we show the unnormalized 
radial functions in Fig. 1 and Fig. 2. 

Furthermore, if the angular momentum $\ell'$ for the first 
excited state is different from the angular momentum $\ell$ 
for the ground state, equation (16) and the constraint (17) become
$$\begin{array}{l}
\beta=4 \alpha \sqrt{a}/
[\ell'(\ell'+1)-\ell(\ell+1)-4(b+6\sqrt{c})/\sqrt{c}], \\
\gamma=4 \alpha \sqrt{c}/[\ell'(\ell'+1)-\ell(\ell+1)], \\
\left[\ell'(\ell'+1)-\ell(\ell+1)-2(b+4\sqrt{c})/\sqrt{c}\right]
/(32\sqrt{ac}) \\
~~~~=\left[\ell'(\ell'+1)-\ell(\ell+1)-4(b+6\sqrt{c})/\sqrt{c}\right]^{-1}
+\left[\ell'(\ell'+1)-\ell(\ell+1)\right]^{-1}. 
\end{array} \eqno (21) $$ 

Setting $a=1.0$, $\ell=0$ and $\ell'=1$, we obtain
$$\begin{array}{llll}
b=-4.2011,~~~~~& c=0.75878,~~~~~& \kappa_{0}=-0.91144, 
~~~~~&\kappa_{1}=1.08856,\\ 
\beta=-1.47683 ~\alpha  , &\gamma=1.74216~\alpha  , 
&E_{0}=-0.82288, &E_{1}=7.17713. \end{array}  \eqno (22) $$

\vspace{4mm}

\begin{center}
{\large 3. Solutions in two dimensions}
\end{center}

For the Schr\"{o}dinger equation in two dimensions with the potential, 
$$V(\rho)=a\rho^{2}+b\rho^{-4}+c\rho^{-6},~~~~~a>0,~~c>0, 
\eqno (23) $$

\noindent
let
$$\psi(\rho, \varphi)=\rho^{-1/2} R_{m}(\rho) e^{\pm im\varphi}, 
~~~~~m=0,1,2,\cdots , \eqno (24) $$  

\noindent
where the radial function $R_{m}(\rho)$ satisfies the 
radial equation
$$\displaystyle {d^{2} R_{m}(\rho) \over d\rho^{2} }
+\left[E-V(r)-\displaystyle {m^{2}-1/4 \over r^{2}} \right] 
R_{m}(\rho)=0. \eqno (25) $$

Making the ansatz for the radial functions of the ground state
and the first excited state:
$$\begin{array}{l}
R_{m}^{(0)}(\rho)=N_{0}\rho^{\kappa_{0}}
\exp [-(\sqrt{a}\rho^{2}+\sqrt{c} \rho^{-2})/2] , \\
R_{m}^{(1)}(\rho)=N_{1}\rho^{\kappa_{1}}
\left(\alpha+\rho^{2}+\gamma \rho^{-2}\right)
\exp [-(\sqrt{a}\rho^{2}+\sqrt{c} \rho^{-2})/2] , 
\end{array} \eqno (26) $$

\noindent
where $\gamma \neq 0$, and substituting Eq. (26) into Eq. (25), we have
$$\begin{array}{l}
\beta [E-\sqrt{a}(2\kappa+5)]=0, \\
\alpha [E-\sqrt{a}(2\kappa+1)]
=\beta[m^{2}-1/4+2\sqrt{ac}-\kappa^{2}-3\kappa-2],\\
\alpha [m^{2}-1/4+2\sqrt{ac}-\kappa^{2}+\kappa]=
\beta [-b+\sqrt{c}(2\kappa+1)]+\gamma[E-\sqrt{a}(2\kappa-3)], \\
\alpha [b-\sqrt{c}(2\kappa-3)]=
-\gamma [m^{2}-1/4+2\sqrt{ac}-\kappa^{2}+5\kappa-6], \\
\gamma [b-\sqrt{c} (2\kappa-7)]=0. \end{array}  \eqno (27) $$

Hence, if the angular momentum $m$ of the ground state is 
the same as that of the first excited state, we obtain from Eq. (27)
$$\begin{array}{ll}
\left(2\sqrt{c}+b\right)^{2}=4c\left[m^{2}+2\sqrt{ac}\right], 
~~~~~&b=-6\sqrt{c}, \\
\kappa_{0}=(3\sqrt{c}+b)/(2\sqrt{c}),~~~~~~~~~~~
&E_{0}=\sqrt{a/c}~(b+4\sqrt{c}), \\ 
\kappa_{1}=(7\sqrt{c}+b)/(2\sqrt{c}),~~~~~~
&E_{1}=\sqrt{a/c}~(b+12\sqrt{c}), \\
\alpha=0, & \gamma=-\sqrt{c}, \end{array} \eqno (28) $$

If $m=0$ and $a=1.0$, the values of the corresponding parameters
are
$$\begin{array}{llll}
b=-12,~~~~~&c=4,~~~~~&\gamma=-2,~~~~~& \\
\kappa_{0}=-1.5,~~~~~ &\kappa_{1}=0.5, ~~~~~& E_{0}=-2,~~~~~
 &E_{1}=6,  \end{array} \eqno (29) $$

\noindent
The unnormalized radial functions are shown in Fig. 3 and Fig. 4.

To summarize, we discuss the ground state and the first 
excited state for the Schr\"{o}dinger equation with the 
potential $V(r)=a r^2+b r^{-4}+c r^{-6}$ using a simple 
${\it ansatz}$ for the eigenfunctions. Two constraints on the
parameters in the potential are arrived at from the compared 
equations. This simple and intuitive method can be 
generalized to the other potentials, such as the sextic 
potential, the octic potential, and the inverse potential.

\vspace{10mm}
{\bf Acknowledgments}. This work was supported by the National
Natural Science Foundation of China and Grant No. LWTZ-1298 from
the Chinese Academy of Sciences. 

%\newpage
%\vspace{5mm}

\end{document}